# Quench dynamics in strongly coupled laser cavities

Mathias Marconi,[1] Julien Javaloyes,[2] Philippe Hamel,[1,*] Fabrice Raineri,[1] Ariel Levenson,[1] and Alejandro M. Yacomotti[1,†]

[1]*Centre de Nanosciences et de Nanotechnologies, CNRS, Université Paris-Sud, Université Paris-Saclay, C2N-Marcoussis, 91460 Marcoussis, France*
[2]*Departament de Física, Universitat de les illes Baleares, C/ Valldemossa km 7.5, 07122 Mallorca, Spain*
(Dated: July 4, 2017)

Strongly coupled dissipative optical cavities with nonlinear interactions give new opportunities to explore symmetry breaking phenomena and phase transitions [1–6], Josephson dynamics [7–11] and quantum criticality [12, 13]. Among the different experimental realizations, photonic crystal coupled nanocavities operating in the laser regime are outstanding systems since nonlinearity, gain/dissipation and intercavity coupling can be judiciously tailored [2, 14, 15]. Yet, although most common scenarios emerge from quasi-dynamical equilibrium where the gain nearly compensates for losses [2, 16], little is known about far out-of-equilibrium dynamics resulting, for instance, from short pulsed pumping inducing a classical "quench". Here we show that bimodal nanolasers generically display transient dynamics after quench which, when projected onto the nonlasing mode, exhibit superthermal light. Such a mechanism is akin to the fast cooling of a suspension of Brownian particles under gravity, with the intracavity intensity playing the role of the inverse temperature. We implement a simple experimental technique to access the probability density functions, that enabled quantifying the distance from thermal equilibrium –and hence the degree of residual order– via the Gibbs entropy. This allowed us to further detect mixing of thermal states with coherent broken parity phases, thus paving the road for investigating far nonequilibrium thermodynamics with multimode optical oscillators.

Besides steady-state dynamical equilibrium and phase transitions, the study of transients after a rapid variation of a parameter (or "quench") has recently captured significant attention in the context of many body systems [17–19]. It has been pointed out that such nonequilibrium dynamics can not only be used as signatures of phase transitions even in presence of dissipation, but also display a wealth of new phenomena in far out-of-equilibrium dissipative systems [18]. This is particularly true in the Mott insulator-to-superfluid phase transitions in optical cavity arrays, in which transients after short pulses give a handle to detect and unveil a rich underlying phase diagram. In open systems photons evolve in a complex way as a result of the subtle interplay between hopping, interaction and losses, analogously to a quantum quench. Surprisingly, the "classical" counterpart of a quenching phenomenon in dissipative cavity arrays remains unexplored. In this article we show that a classical quench in a strongly coupled laser cavity system following short pump pulses is akin to a fast cooling process in an ensemble of Brownian particles under gravity: transients generically display long-tailed statistical distributions as signatures of far nonequilibrium dynamics.

Brownian particles are mesoscopic objects submitted to external forces but also in contact with a thermal bath, e.g. pollen submerged into a fluid as in Robert Brown experiments [20]. The interactions with the thermal reservoir provides at the same time dissipation and fluctuations that lead to thermal equilibrium. In the high friction limit, the altitude $z_j$ of the particle $j$ follows the Langevin equation of motion $\gamma dz_j/dt = -g + \sqrt{2\gamma k_B T/m}\xi_j(t)$, where $\gamma$ is the viscous damping rate, $g$ is gravitational acceleration, $k_B$ the Boltzmann constant, $m$ the mass, $T$ the reservoir temperature and $\xi_j(t)$ a Gaussian white noise. We consider a hard (reflecting) wall at $z = 0$ as the boundary condition and re-scale the variables as $T \to T/\theta_g$ with the characteristic temperature $\theta_g = g^2 m/\gamma^2 k_B$, $z \to z\gamma^2/g$ and $t \to \gamma t$. The normalized Langevin equation reads

$$\frac{dz_j}{dt} = f(z) + g(z)\xi_j(t), \qquad (1)$$

where $f(z) = -1$ and $g^2(z) = 2T$ are the normalized drift and diffusion coefficients, respectively (we assume a generic dependence with the coordinate $z$, null in this case). Equation 1 has the following associated Fokker-Planck equation governing the evolution of the probability distribution $\rho(z)$ for $z > 0$,

$$\frac{\partial \rho}{\partial t} + \frac{\partial}{\partial z}(f\rho) = \frac{1}{2}\frac{\partial^2}{\partial z^2}(g^2 \rho). \qquad (2)$$

The equilibrium solution of Eq. 2 is simply the exponential distribution of the potential energy $U(z) = z$ (see Figs. 1.a and c),

$$\rho_{eq}(z, T) = \mathcal{N}e^{-\beta z}, \qquad (3)$$

---
* Present adress: Cabinet Camus-Lebkiri, Immeuble Stratege A, 51 rue Ampere, 31670 Toulouse
† Alejandro.Giacomotti@u-psud.fr

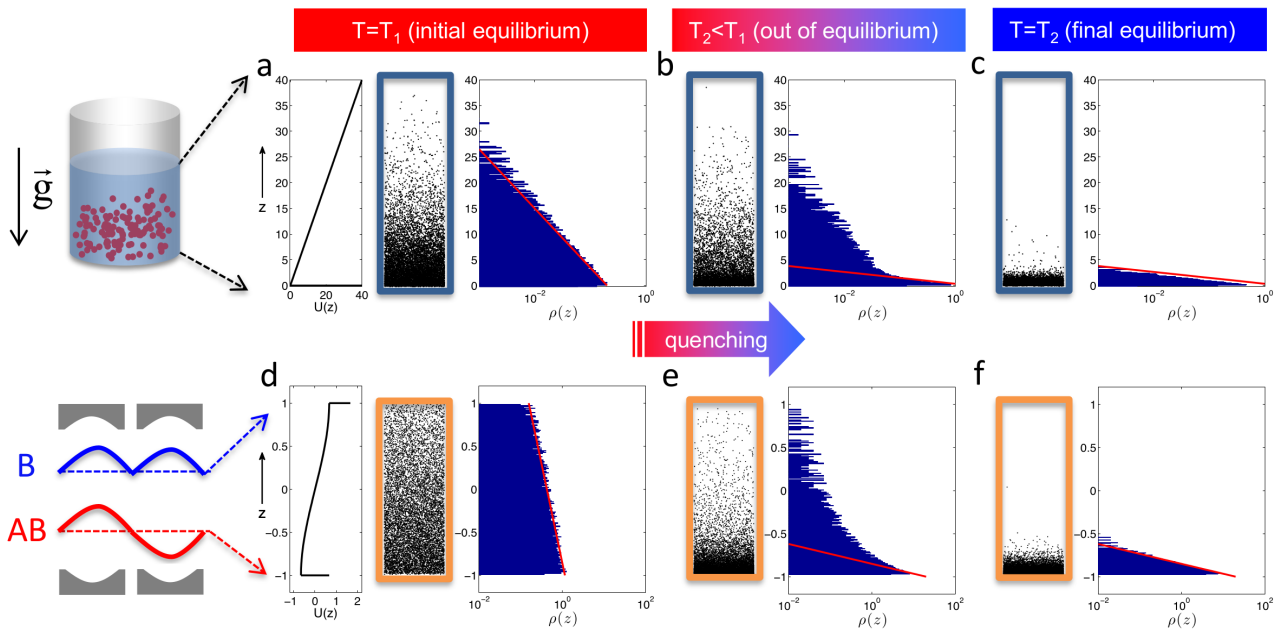

Figure 1. Quenching analogy between a suspension of Brownian particles in a fluid (a-c) and a double cavity laser with mode energy flow from bonding (B) to anti-bonding (AB) modes (d-f). a,d) Initial equilibrium. From left to right: potential energy, a snapshot of the physical ensemble ($10^4$ particles), height histogram (blue bars) and the steady state equilibrium distribution $\rho_{eq}(z,T)$ (red line, Eq. 3). b,e) Quenched phase. The far nonequilibrium distributions are due to the abrupt decrease of the reservoir temperature from $T_1$ to $T_2 < T_1$. Particles located at high altitudes remain "hot" while lower particles cool down more rapidly. c-f) Final equilibrium at $T = T_2$.

with $\beta = 1/T$ and the normalization constant is $\mathcal{N} = \beta$. Equation 3 exactly reproduces the Boltzmann distribution of an "ideal gas" in presence of gravity, $n(z) = (mg/KT)(N/V)\exp(-mgz)$ in its dimensional version, which does not depend on viscosity.

If the system is initially at thermal equilibrium with temperature $T = T_1$, two widely different paths, or *cooling processes*, are possible towards a temperature $T_2 < T_1$. First, the reservoir temperature $T(t)$ may decrease slowly as compared to the characteristic timescale of the particle $\gamma^{-1}$. In our re-scaled units this means $\dot{T}/T \ll 1$. In this "adiabatic" case, the distribution during the cooling process is defined by the equilibrium distribution with a time-dependent temperature $\rho(z,t) = \rho_{eq}[z,T(t)]$ (see movie S1, Supplementary Material).

Another possibility consists in decreasing the temperature abruptly. The system will remain in an out-of-equilibrium state for some time, with a statistical distribution that will not correspond to the exponential equilibrium (see Fig. 1.b and movie S2, Supplementary Material). In such a transient phase after quench the particles are simply falling and the Brownian fluctuations can be neglected, hence the dynamics is dominated by transport (drift) rather than diffusion. Collisions with the bottom wall make particles thermalize faster: those with low potential energy become cold, whereas high particles remain hot. Here we show that this simple picture unveils quench phenomena in more complex systems such as strongly coupled dissipative optical cavities operating in a laser regime (Fig. 1.d).

The analogy between falling Brownian particles and compound cavity laser systems relies on three important features: i) coupled cavities may support several (here two) eigenmodes which, as long as the cavities are filled with active (gain) media, interact through mode-mode scattering and/or gain competition leading to energy flow (drift) among the modes [16, 21]; ii) stochastic fluctuations are intrinsic because of spontaneous emission noise; and iii) the equilibrium below laser threshold is thermal. We define $I_B$ the symmetric (hereby "bonding", B) mode intensity of the photonic molecule, $I_{AB}$ the anti-symmetric ("anti-bonding", AB) one, and the population imbalance $z = (I_B - I_{AB})/(I_B + I_{AB})$, in such a way that $z = 1$ for the B and $z = -1$ for the AB mode ($z \in [-1, 1]$). Let us assume that the optical losses of the AB mode ($\kappa_{AB}$), are lower than those of the B mode ($\kappa_B$) by a factor $\gamma = (\kappa_{AB} - \kappa_B)/2 < 0$, hence AB will be the stable lasing mode. In absence of noise $z \to -1$ as $t \to \infty$, which is a consequence of the drift force, analogously to the gravity for the Brownian particles. If we now consider spontaneous emission (SE) fluctuations, and since B is a nonlasing mode, the probability distribution of its intensity $I_B = (z+1)I/2$, where $I = I_B + I_{AB}$ is the total intensity, is likely to be exponential. As long as the total intensity fluctuations $\langle(\delta I)^2\rangle$ are negligible with respect to mode-exchange energy fluctuations

$\langle(\delta z)^2\rangle$, the equilibrium distribution of $z$ is also expected to be exponential with a maximum at $z = -1$ (Figs. 1.d and f). This can be demonstrated using a semiclassical model for two coupled semiconductor laser cavities leading to a Langevin equation of the form given in Eq. 1 (see SI), but this time $f(z) = \gamma_z(1-z^2) - S_p I^{-1} z$ and $g(z) = \sqrt{(1-z^2)S_p I^{-1}}$, where $\gamma_z$ is the effective damping ($\gamma_z \approx 2\gamma$ close enough to the laser threshold) and $S_p$ a normalized spontaneous emission rate in the cavity. Diffusion is now given by a multiplicative noise term, and the steady state equilibrium is exponential (Eq. 3) as expected, with $\beta = 2\gamma_z I S_p^{-1}$ and $\mathcal{N} = \beta/(2\sinh\beta)$, in such a way that $\beta^{-1} \sim S_p I^{-1}$ can be identified as the "effective temperature of the bath". Suddenly increasing the intensity is thus equivalent to decreasing the reservoir temperature. Such a rapid cooling leads to a quench and a subsequent nonequilibrium distribution depicted in Fig. 1e, before reaching the final thermal equilibrium in Fig. 1f. The far-from-equilibrium phase shows a long tailed, "multi-exponential" distribution which, when projected onto the nonlasing (B) mode is likely to produce superthermal light, namely with a degree of second order coherence $g_B^{(2)} \equiv g_B^{(2)}(\tau = 0) = \langle I_B^2\rangle/\langle I_B\rangle^2 > 2$, $g_B^{(2)} = 2$ being the thermal limit.

In order to experimentally show super-thermal, long-tailed Probability Density Function (PDF) transients after quench in a photonic system, we use two coupled nanolasers under short pulsed pumping, which is a natural path to experimentally access quench dynamics. Two coupled photonic crystal cavities are optically pumped at the geometrical center with short (100 ps-duration) pulses (Fig. 2a). The emission is collected with a high numerical aperture microscope objective, and its back focal plane is imaged through a lens to obtain the far-field pattern (Fourier plane, Fig. 2a). In order to temporally resolve the pulse energy of the two coupled modes B and AB, we set up two confocal detection paths: two single mode fibers are used as pinholes to simultaneously select two small regions on the Fourier plane and detect their intensities. These regions are located at the center ($k = 0$) for the B-mode detection, and shifted along the $x$ direction for the AB-mode (Fig. 2b). Signals are sent to two identical low noise (200 fW/$\sqrt{Hz}$), 660 MHz-bandwidth avalanche photodiode (APD) detectors. Hence, the detectors integrate the output pulses resulting in peak APD signals proportional to the pulse energy, which is eventually calibrated in photon number ($p$). Typical $p$-time series contain $10^4$ pulses. Figure 2a shows a segment of two simultaneous $p$-time series showing both the AB and B-mode signals. We point out that, since the cavities operate in a laser regime, each output pulse, containing projections on both B and AB modes, results from a single trajectory in phase space, whose time integration leads to single pulse photon number.

Second order correlations of the pulse energies $g_u^{(2)}$ (see SI), or equivalently photon number $g_p^{(2)}$ for both modes, are computed from the variances of the $p$-time traces and depicted in Fig. 2b. Our method is validated by means of standard Hanbury Brown and Twiss experiments (see Fig. 2b) [25, 26]. It is clearly observed that second order correlations measured with both techniques are within experimental errors. These are related to time-integrated autocorrelation functions which can be approximated as snapshots close to the pulse maxi-mum $g^{(2)}(t_{max}, \tau \approx 0)$ (see SI). In particular, $g_p^{(2)}$ for the B-mode reveals superthermal emission $[g_p^{(2)} > 2]$ for $P_{pump} \gtrsim 1.5 P_{th}$, while the AB mode emission remains Poissonian $[g_p^{(2)} \approx 1]$. Second order correlation functions and superthermal light statistics have already been observed in other photonic systems such as pseudothermal light sources [27], superradiant quantum dots [28], or bimodal micropillar with quantum dots, where superthermal light due to mode switching instabilites has been reported [29, 30]. The mechanism for superthermal light generation in our case, instead, is a far non-equilibrium quenching process, which can be demonstrated on the basis of the full probability distributions.

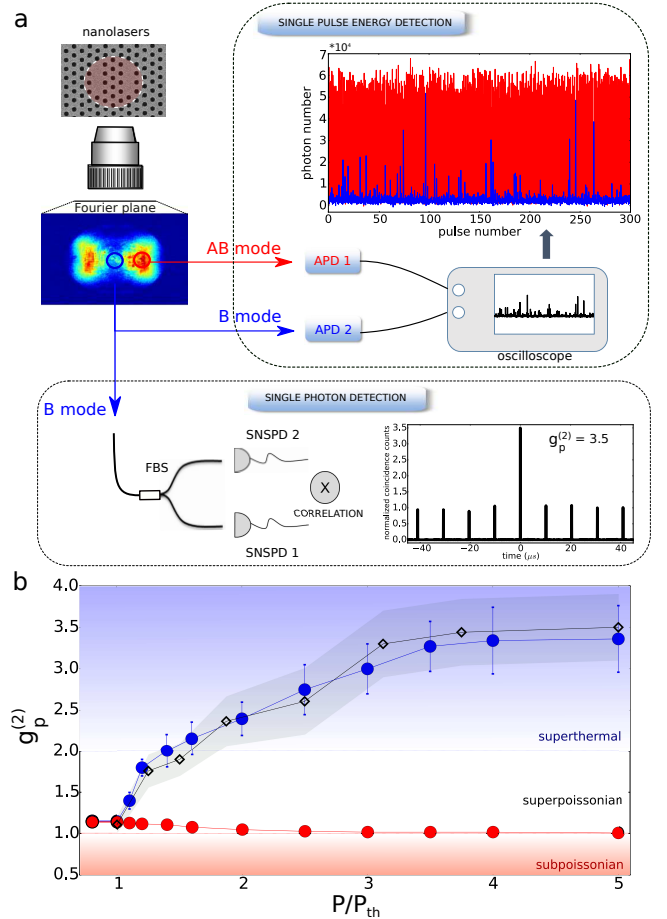

Figure 2. Experimental mode-resolved photon number. a) "Single pulse energy detection" scheme (top): pulse-to-pulse measurement technique of emitted photon numbers. Simultaneous photon collection from the two modes and further detection with APDs allows us to obtain a sequence of pulses ($p$-time trace) from which PDFs and $g_p^{(2)}$ values are computed. "Single photon detection" scheme (bottom): light from the bonding mode is also sent to a HBT setup to provide direct measurement of second order coherence, used as a cross-check. SNSPD: superconducting nanowire single photon detector. Inset: normalized coincidence count histogram of the B-mode for $P = 4P_{th}$. b) $g_p^{(2)}$ for B and AB modes at different pump powers computed from the variance of the $p$-time traces shown in a). Black diamonds show the corresponding HBT autocorrelation measurements: symbols represent the area under the central peak of the autocorrelation normalized to the averaged area under secondary peaks (error bars in grey shadow), in good agreement with $g_p^{(2)}$ measurements (see SI for a theoretical comparison).

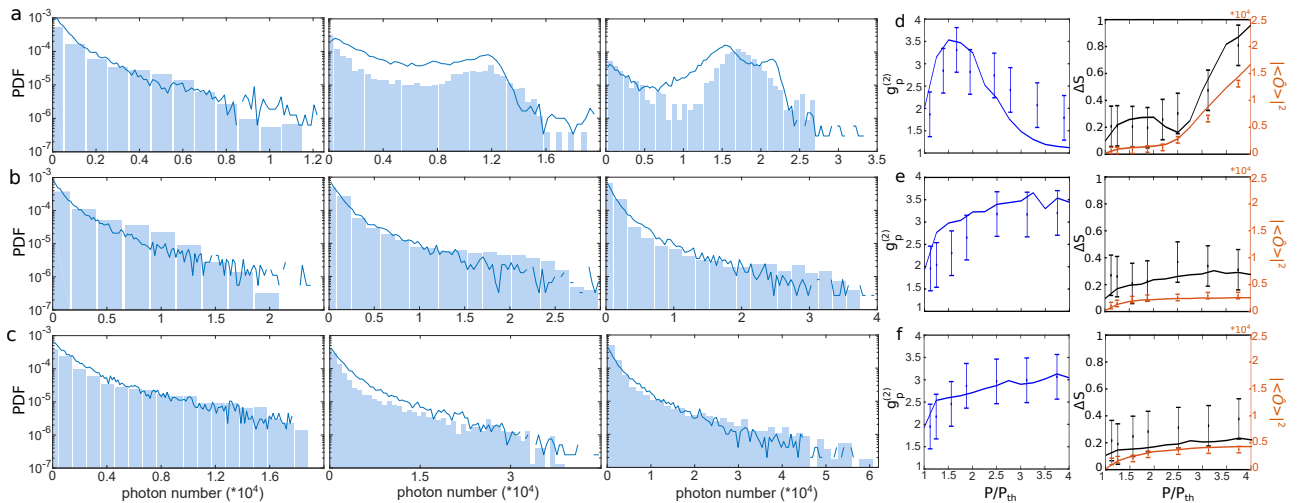

Figure 3. Statistical distributions of photon number for the B-mode. a-c) Experimental (bars) and theoretical (lines) PDFs for increasing pump powers (left : $P = 1.5 P_{th}$, middle : $P = 2 P_{th}$, right: $P = 3 P_{th}$) and different intercavity coupling levels: a) $K\tau = 3.3$, b) $K\tau = 6$ and c) $K\tau = 10$. d-f) Experimental (symbols) and theoretical (lines) degree of second order coherence (left), entropy production (right, black) and order parameter (right, orange) as a function of $P$ for d) $K\tau = 3.3$, e) $K\tau = 6$ and f) $K\tau = 10$.

Figures 3a-c show experimental and numerical photon number PDFs for three strong intercavity coupling parameters [15], $K\tau = 3.3$, $K\tau = 6$ and $K\tau = 10$ respectively, and three different pump rates, $P_{pump} = 1.5 P_{th}$, $P_{pump} = 2 P_{th}$ and $P_{pump} = 3 P_{th}$ from left to right. For the two larger coupling parameters ($K\tau = 6$ and $K\tau = 10$) the PDFs evolve from nearly exponential to long-tailed PDFs as the pump power is increased, showing very good agreement between experimental data and numerical simulations of a stochastic mean-field model. As for second order moments, the measured PDF can also be interpreted as snapshots close to pulse maximum $t = t_{max}$: stronger pump induces more intense laser output pulses, and hence a stronger quench, which leads to more pronounced long-tailed distributions. In the following we will investigate in detail the time dependent quench dynamics upon short pump pulses.

We use mean-field dynamical equations in presence of noise to model the response of the system under short pulse pumping. We label the left and right coupled nano-cavities with indexes $j = 1$ and $j = 2$, and define $A_j = \sqrt{2} R_j \exp(i\psi_j)$ the normalized slowly varying complex field amplitude and $D_{1,2} = N \pm n$ the in-site normalized population inversions. As such, $N$ and $n$ represent the average and the population difference between the two carrier reservoirs. We deal with a transient, far nonequilibrium situation, as a consequence of a short incoherent pump pulse. A particular solution takes the form of a trajectory in phase space, that eventually comes back to the neighborhood of the rest equilibrium point ($|A_j|^2 = 0$). The stochastic fluctuations coming from SE noise, which are inherent of micro/nanolasers [22–24], have a main impact: the trajectory in phase space triggered by a given pump pulse will start from random initial phases. A useful representation is the Bloch sphere, where $\theta = 2 \arctan(R_1/R_2) \in [0, \pi]$ is the polar angle, which is a measure of the photon density imbalance between the two cavities, and the azimuthal angle $\Phi = \psi_1 - \psi_2$ is the phase difference between the sites. In this representation, the bonding and anti-bonding modes correspond to two opposite points over the equator, $(\theta, \Phi) = (\pi/2, 0)$ and $(\theta, \Phi) = (\pi/2, \pi)$ (see Fig.4).

A most important step in out theoretical analysis is to assume large mode frequency splitting $2K$ and small threshold difference $\gamma$ between the bonding and the anti-bonding modes, compatible with the coupled nanolaser experiment. This enables a separation of variables in which the total intensity and carrier density can be computed independently, $[I(t), N(t)]$ (Fig. 4a), and become a parametric forcing of the Bloch sphere and carrier population difference dynamics. Furthermore, a reduction to a 1D stochastic differential equation of the fractional population imbalance $z = \sin\theta\cos\Phi$ is possible: the slowly evolving $z$ variable is indeed a con-venient adiabatic invariant to describe energy transfer between the modes (see SI), leading to Eq. 1 and the subsequent Fokker-Planck equation 2 with a diffusion parameter $g^2 \sim S_p I(t)^{-1}$ and effective damping $\gamma_z(t) = 2\gamma - \alpha NI (2TK)^{-1}$ ($\alpha$ is the phase-amplitude coupling factor of the semiconductor laser and T is the ratio between the carrier to the cavity lifetimes). Hence, the steep intensity increase (Fig. 4a) –with buildup timescale corresponding to the cavity photon lifetime, $\dot{I}(t)/I(t) \sim 1$– is synonym of a cooling down process much faster than the typical time scale of the mode energy exchange $\gamma_z^{-1} \sim 10$.

We consider repetitive sequential cycles of optical pumping such that each response to a single pump pulse is a particular statistical realization, so that we can follow an ensemble of trajectories in parallel ($10^4$). The statistical distributions on the Bloch sphere are shown for different elapsed times before (Fig. 4b,right) and within (Fig. 4c-d,right) the $I(t)$ laser pulse. We can observe two distinct situations: a quasi-equilibrium random phase distribution below threshold (Fig. 4b,right), and non-equilibrium (Fig. 4c-d,right), which become peaked around the antibonding fixed-point $(\theta,\phi) = (\pi/2,\pi)$ at the $I(t)$ maximum (Fig. 4d,right). Figures 4b-d (left pannels) show the 1D PDFs of population imbalance $z$. While the probability distribution is initially at the exponential equilibrium corresponding to the high temperature $\beta = \beta_0$ (Fig. 4b), it is forced to evolve in a basically noiseless situation ($\beta \to \infty$). In this initial phase of relaxation towards a much lower "reservoir" temperature, nonlinear transport dominates over diffusion in Eq. (2) and we can solve it setting $S_p = 0$. We find that the solution of the transport problem with initial condition $\rho_t(z,t_0) = \rho_0(z)$ reads

$$\rho_t(z,t) = \frac{4e^\tau \rho_0 \left[\frac{1+z+e^\tau(z-1)}{1+z-e^\tau(z-1)}\right]}{[1+e^\tau + z(1-e^\tau)]^2} \quad (4)$$

where we defined the effective normalized time $\tau(t) = \int_{t_0}^t 2\gamma_z(s)\,ds$ that takes into account the temporal dependence of $\gamma_z(t)$ via the intensity carrier product $NI(t)$. Therefore $\tau$ may increase faster than $t$ in the time interval around the pulse where $\alpha NI > 4\gamma TK\alpha^{-1}$. Equation 4 can thus be regarded as an instantaneous reservoir cooling down or "free falling" approximation of the pulsed-pump nonequilibrium distribution where the only fitting parameter is $t_0$ (see Figs. 4b-d,left, green lines).

We characterize these out-of-equilibrium states by first computing the time dependent second order coherence of the bonding mode at vanishing time delay, $g^{(2)}(t,0)$, plotted in Fig. 4a. Before and after the laser pulse, i.e. for $I(t) \approx 0$, $g^{(2)}(t,0) \approx 2$, which corresponds to thermal emission. Close to the pulse maximum ($t_{max}$) the B mode becomes superthermal, $g^{(2)}(t=t_{max},0) \approx 3$. After-pulse superbunching can also be observed, $g^{(2)}(t=t_2,0) \approx 8$.

Our last step is use the information of the full statistical distributions –also experimentally accessible– to quantify deviations from equilibrium by means of the en-tropy $S(t) = -\int \rho(z,t) \ln[\rho(z,t)]dz$. Now, in order to measure the distance from thermal equilibrium we consider a subsequent irreversible transformation from the out-of- equilibrium state $\rho(z,t)$ as the initial state, and we let the system evolve without net energy exchange with the reservoir. Namely, we define the final state as an equilibrium state of temperature $T_f$ witch results from inverting $U_{eq}(T_f) = U(t)$, where $U(t) = \int z\rho(z,t)dz$ is a time-dependent internal energy, leading to a function $T_f[U(t)]$. As a result, the exchanged entropy with the reservoir vanishes and the only source of entropy production is the degree of internal order; we thus define $\Delta S(t)$ as

$$\Delta S(t) = S_{eq}\{T_f[U(t)]\} + \int \rho(z,t)\ln[\rho(z,t)]dz. \quad (5)$$

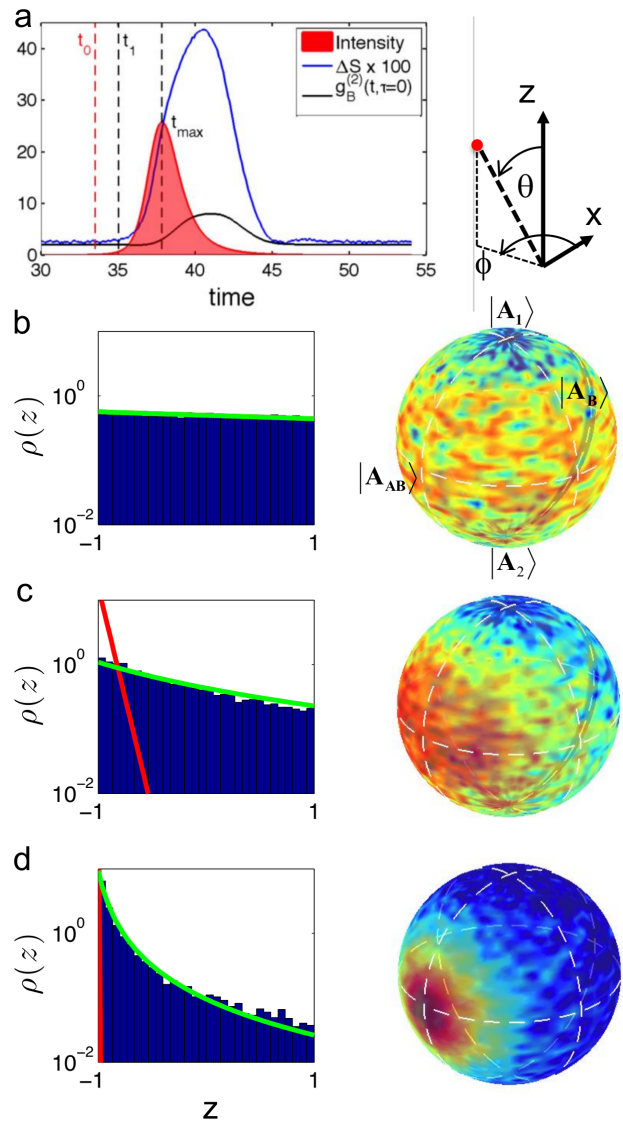

Figure 4. Numerically simulated response of coupled nanolasers system under short pulse pumping (centered at $t_{pump} = 32$). a) Evolution of total laser output intensity (red), entropy (blue) and second order autocorrelation for the bonding mode (black) . b-d) Snapshots of the PDFs for the fractional population imbalance (left) and Bloch sphere representations (right) at b) $t = 20$ (before laser build-up), c) $t_1 = 35$ (during the early phase build-up) and d) $t_{max} = 37.8$ (pulse maximum). Red curves on PDFs represent the equilibrium solutions of Eq. 2 to a final temperature as given by the instantaneous intensity. Green curves are $T = 0$, "free falling" approximations of the PDF (Eq. 4) with fitted $t_0 = 33.5$. Simulation parameters: $K = 7$, $\gamma = -0.1$ and $P = 4P_{th}$.

Note that this transformation can be thought as a relaxation to equilibrium with conserved internal energy, as if the system was "thermally isolated". As such, it quantifies the degree of internal residual order –or coherence– of the system, which is maximum for the farthest nonequilibrium state (see SI for more details). Figure 4a shows $\Delta S(t)$ which is significantly high close to the pulse maximum, $\Delta S(t_{max}) \approx 0.25$, and can be considered as a measure of the Brownian particle-like quench.

Now, a link can be established between the time dependent evolutions of the statistical observables after quench and the experimental results obtained with the single pulse energy detection scheme. The superthermal quenches for $K\tau = 6$ and $K\tau = 10$ asymptotically reach $g_p^{(2)}$ values $\sim 3-3.5$ (Figs. 3e-f, left), in agreement with $g^{(2)}(t_{max}, 0) \approx 3$ in Fig. 4a. The Brownian particle character of the quench is confirmed by computing the entropy $\Delta S$ as defined in Eq. 5. The monotonic entropy increase towards $\Delta S \sim 0.2 - 0.3$ is compatible with transients after quench for $K\tau = 6$ and $K\tau = 10$, and correspond to $\Delta S(t_{max})$ predicted by the model (see Fig. 4a). However, neither $g_p^{(2)}$ nor the entropy have the same behavior as the pump power is increased for $K\tau = 3.3$. On the one hand, it can be observed that $g_p^{(2)}$ tends to one for increasing pumping, which is an indication of the presence of a coherent phase. On the other hand, local maxima in the distributions for $K\tau = 3.3$ (Figure. 3a) lead to larger deviations from equilibrium which, even in good agreement with the full numerical simulations of the stochastic mean-field model, are not predicted by our simple Brownian particle description. Instead, the large entropy departure for $K\tau = 3.3$ and $P > 2.5P_{th}$ (Figure. 3d,right) can be interpreted as a consequence of symmetry breaking phenomena taking place for such coupling parameters, as we have recently explored in continuous wave operation [2]. As proposed in Ref. [4], a symmetry breaking phase transition can be detected from the evolution of an order parameter, $|\hat{O}| = |A_1 + A_2|$, which will grow up under symmetry breaking conditions. We observe in Figs. 3d-f (right panels) that $\langle|\hat{O}|^2\rangle$ remains small for $K \geq 6$, while it experiences an abrupt increase for $K = 3.3$ and $P > 2.5P_{th}$, which constitutes an experimental evidence –in the sense of a quench– of a symmetry breaking phase transition. Hence, $K = 3.3$ with $P > 2.5P_{th}$ corresponds to the mixing of a thermal state with a symmetry broken coherent phase, which is beyond the Brownian particle picture. It shows the increasing statistical complexity of the dissipative coupled cavity system when approaching the highly nonlinear limit in which the nonlinear laser frequency shift overcomes the tunnel energy $K$ [2].

In conclusion we have shown that a bimodal nanolaser under pulsed pumping behaves as a thermodynamic ensemble of small Brownian particles subjected to both: i) a drift force provided by mode competition, and ii) diffusion in a thermal bath given by spontaneous emission noise. Both stochastic fluctuations and dissipation lead to a Boltzmann equilibrium for the fractional population imbalance of the mode energies. In this framework the inverse total laser intensity behaves as the reservoir temperature, such that a pulsed pumping configuration naturally induces a quench. As a result, far nonequilibrium transients emerge in the nanolaser system showing long tailed superthermal distributions of the nonlasing mode, which has been experimentally demonstrated by means of a single pulse energy detection scheme. These results open up a new paradigm to investigate classical phase transitions and out-of-equilibrium thermodynamics using multimode nonlinear dissipative optical cavities.

---